\documentclass{article}
\usepackage[utf8]{inputenc}
\usepackage[left=1in, right=1in, top=1in,bottom=1in]{geometry} 
\title{Theory In, Theory Out: The uses of social theory in machine learning for social science}
\author{
  Jason Radford\thanks{Both authors contributed equally to this manuscript}\\
  \texttt{j.radford@northeastern.edu}\\
  Network Science Institute\\
  Northeastern University
  \and
  Kenneth Joseph\textsuperscript{*}\\
  \texttt{kjoseph@buffalo.edu}\\
  Computer Science and Engineering\\
  University at Buffalo
}

\date{}

\usepackage{graphicx}

\begin{document}

\maketitle
\begin{abstract}
Research at the intersection of machine learning and the social sciences has provided critical new insights into social behavior.  At the same time, a variety of critiques have been raised ranging from technical issues with the data used and features constructed, problematic assumptions built into models, their limited interpretability, and their contribution to bias and inequality. We argue such issues arise primarily because of the lack of social theory at various stages of the model building and analysis. In the first half of this paper, we walk through how social theory can be used to answer the basic methodological and interpretive questions that arise at each stage of the machine learning pipeline. In the second half, we show how theory can be used to assess and compare the quality of different social learning models, including interpreting, generalizing, and assessing the fairness of models. We believe this paper can act as a guide for computer and social scientists alike to navigate the substantive questions involved in applying the tools of machine learning to social data.
\end{abstract}

\section{Introduction}

Machine learning is increasingly being applied to vast quantities of social data generated from and about people \cite{lazer_computational_2009-1}.  Much of this work has been fruitful.  For example, research using machine learning approaches on large social datasets has allowed us to provide accurate forecasts of state-level polls in U.S. elections \cite{beauchamp_predicting_2017}, study character development in novels \cite{bamman_bayesian_2014}, and to better understand the structure and demographics of city neighborhoods \cite{cranshaw_livehoods_2012,hipp_measuring_2012}.  The increasing application of machine learning to social data has thus seen important success stories advancing our understanding of the social world.

At the same time, many social scientists have noted fundamental problems with a range of research that uses machine learning on social data \cite{lazer_data_2017,crawford_ai_2019}.  Machine learning on social data often does not account for myriad biases that arise during the analysis pipeline that can undercut the validity of study claims \cite{olteanu_social_2016}. Attempts to identify criminality \cite{wu_automated_2016} and sexuality \cite{wang_deep_2018} from people’s faces and predicting recidivism using criminal justice records \cite{larson_how_2016} have led to critiques that current attempts to apply machine learning to social data represent a new form of physiognomy \cite{arcas_physiognomys_2017}. Physiognomy was the attempt to explain human behavior through body types and was characterized by poor theory and sloppy measurement \cite{gould1996mismeasure}. It ultimately served to merely re-enforce the racial, gender, and class privileges of scientists and other elites. Today it is considered pseudoscience. 

Acknowledging these misappropriations of machine learning on social data, researchers have largely sought out technical solutions, both old and new.  For example, in response to claims that algorithms embedded in policy decisions often provide unfair advantages and disadvantages across social groups, scholars in the Fairness, Accountability and Transparency (FAT*) community have proposed new algorithms to make decisions more fair. Similarly, researchers in natural language processing have proposed several new methods to ``de-bias'' word embeddings' representation of gender, race, and other social identities and statuses. \cite{bolukbasi_man_2016-1}. 

The primary contribution of this paper is to argue and show that \emph{at each step of the machine learning pipeline, problems arise which cannot be solved using a technical solution alone, but with the use of social theory}. Social theory is the set of scientifically-defined constructs like race, gender, social class, inequality, family, and institution and their empirically-identified causes and consequences. Using social theory in machine learning means engaging these constructs as they are defined and described scientifically and accounting for the established mechanisms and patterns of behavior engendered by these constructs. For example, Omi and Winant’s \cite{omi2014racial} racial formation theory argues that race is a social identity that is constantly being constructed by political, economic, and social forces. What makes someone ``Black'' or ``White'' in the United States and the opportunities and inequities associated with this distinction have changed dramatically throughout history and continues to change today. While there are other scientific definitions of race and active debates about its causes and consequences, engaging with them at each stage in the pipeline allows us to answer critical questions about the models we should use, features we should select, and generalizations we can make. 

\begin{figure}
    \centering
    \includegraphics[width=\linewidth]{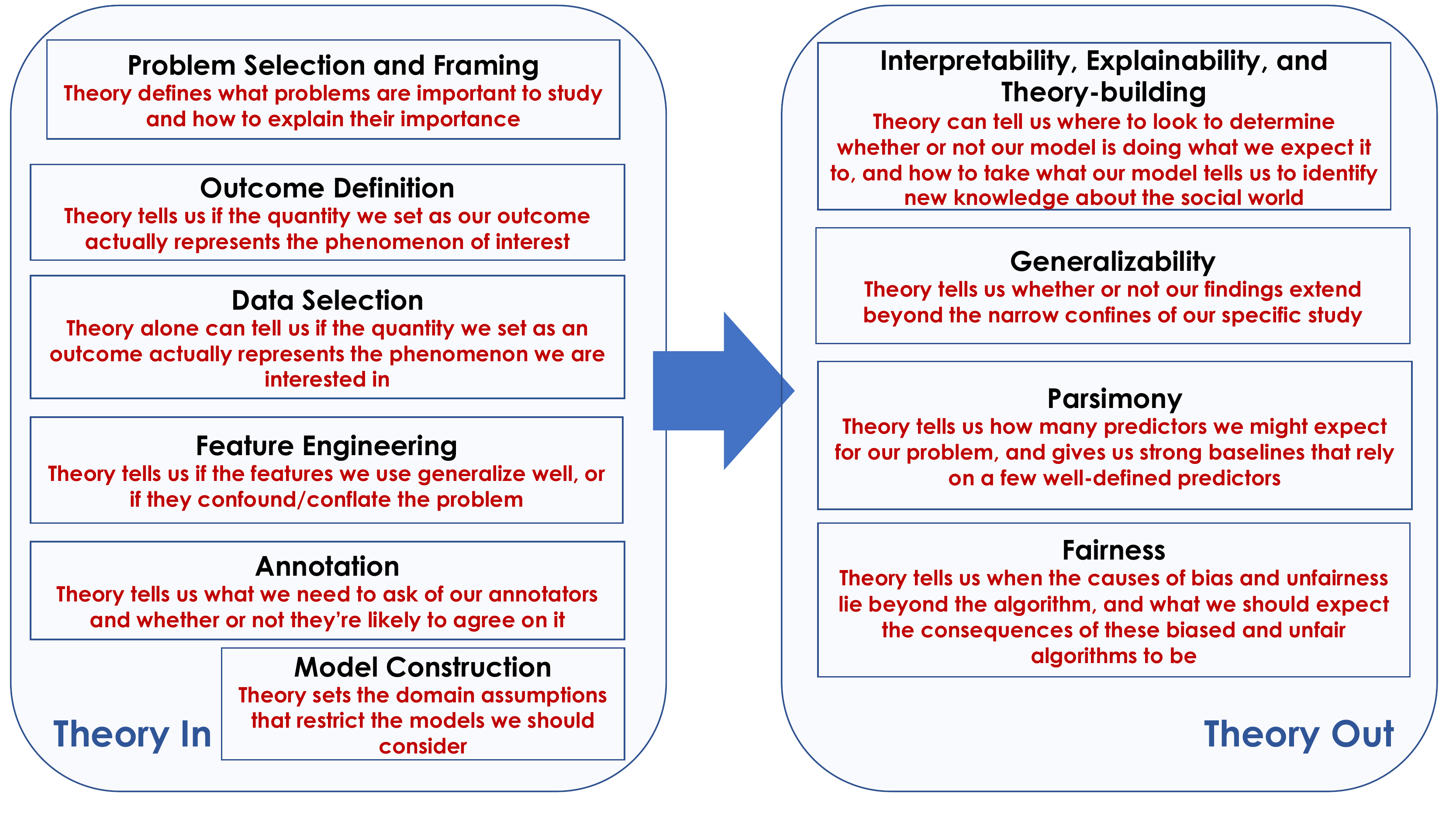}
    \caption{An overview of the Theory In and Theory Out sections of the paper. Each box within Theory In and Theory Out is a subsection. The box gives the name of the subsection and the claim we make about how social theory can be used to improve our technical solutions for addressing that problem.}
    \label{fig:intro}
\end{figure}

In this paper, we explain how social theory helps us solve problems arising at every step in the machine learning for social data pipeline. This is outlined in Figure~\ref{fig:intro}. 

The paper is structured into two broad section. In the \emph{Theory In} section, we discuss how social theory can help us model social data by informing how we identify important problems and determine the right class of models. In the \emph{Theory Out} section, we talk about different desiderata that we have for the models and results we produce, like generalizability, and discuss how social theory can help to improve these outputs of our work.

Both the Theory In and Theory Out sections have multiple subsections. Although each subsection addresses a unique step in the pipeline, solutions problems identified in one step often enable solutions to problems in others.  For example, a model that is parsimonious is often more interpretable.  A lack of a solution to a problem in one step can also prohibit a solution to issues that might arise downstream. At the highest level, a lack of social theory going into the model is likely to stymy drawing theory out. These overlaps are a strength, rather than a weakness, of the structure of this article. Like Olteanu et al. \cite{olteanu_social_2016}, we believe that by emphasizing both the uniqueness and the critical relationships between different pieces of the pipeline, we can understand how failure to address critical problems can propagate from one step to the next.

\section{Related Work}

Social scientists have long established that theory can solve methodological and analytic issues that new techniques cannot. For example, Mario Small has argued that theory alone can address questions of how best to measure what it means for one to have a close social tie \cite{small_someone_2017}. In the present work, we regularly draw on this literature, seeing many parallels between prior methodological innovations like linear models and sequence analysis \cite{abbott_transcending_1988,abbott_sequence_1995}. 

Other scholars working at the intersection of machine learning and social science have also proposed important critiques upon which we draw throughout the paper. These critiques fall into four broad categories. First, Hanna Wallach  and her colleagues have argued that many machine learning papers focus too heavily on prediction relative to explanation \cite{wallach_computational_2018} or measurement \cite{jacobs2019measurement}. The prioritization of prediction leads to models that perform well for unknown reasons, leading to ad hoc justifications for model decisions and performance. It also leads to models which output parameters that cannot be used to directly measure real-world quantities we are interested in. A lack of focus on measurement leads, as we will also discuss, to a failure to acknowledge the always imperfect association between what we are trying to measure and what we are able to quantify. 

Others, like Laura Nelson \cite{nelson_computational_2017}, argue that machine learning applied to social data must come hand-in-hand with the development of new theory. From their perspective, we do not necessarily know what model will work for what data, nor do we typically have theory to tell us what to expect.  Consequently, we need to create new theory as we develop and run methods on our data.  This approach helps us to understand why machine learning models might present results that at first seem unintuitive, but that do reflect genuine patterns that should force us to reconsider our understanding of the social world.  However, it also requires an a priori understanding of the potential theories that could apply, and seeks to adapt this existing theory, rather than create new theory entirely ex post facto.

Still others have taken specific studies or sets of studies to task, arguing that they fail to understand the socio-technical context in which their data are produced \cite{tufekci_big_2014,lazer_computational_2009-1,olteanu_social_2016}.  For example, Tufecki \cite{tufekci_big_2014} argues that, despite generic claims above universal social behavior in many papers, research using Twitter data is unlikely to tell us much about social interaction on Facebook because the two have very different rules, data, and norms.  Similarly, Olteanu et al. \cite{olteanu_social_2016} provide a dizzying array of potential pitfalls in analyzing social data, emphasizing the need for time-tested ideas from the social sciences, like matching and analyses of sample bias, to address these issues. These critiques point to the fact that machine learning is often done with one eye closed to the peculiarities of the data. 

Finally, with the advent of algorithms that are making increasingly high-impact decisions, the FATE community\footnote{http://fatconference.org} has arisen to study how these algorithms can serve to reflect social biases embedded in the complex sociotechnical systems within which they are embedded, and how we might be able to address these issues.  However, recent critiques of the fairness literature argue that far too much emphasis has been placed on technical ``solutions'' to unfair and/or ``biased'' algorithms, relative to the structural causes and consequences of those algorithms \cite{hoffmann_where_2019,green_fair_2018,crawford_ai_2019}. Such scholarship has argued that social science disciplines need be at the forefront of our understanding of how to address these root causes. 

Each of these critiques---that prediction does not equal understanding, that we must be ready to construct new theory to interpret our results, that myriad biases lurk in the data and methods we use, and that these methods can result in discriminatory outcomes with systemic causes--- is critical in pushing us towards a better, more responsible application of machine learning towards social data. Further, the many works reviewed below that apply machine learning to social data with these issues in mind provide a further glimpse into the potential for this version of the science.

In the present work, we seek to unify these critiques, arguing that each of them are levied at different pieces of the same underlying problem---attempts to use technology, or ad hoc, post ex facto reasoning, to address problems only social theory can solve.  We argue below that theory alone can lead us to the valid explanatory models sought by Wallach \cite{wallach_computational_2018}, to ensure we draw correct conclusions from initially unintuitive results, to help us characterize dangerous assumptions in our data collection processes, and to help us understand and address discriminatory, biased, or unfair model behavior.  

\section{Theory In}\label{sec:theory_in} 
In this section, we discuss the pipeline for studies that use machine learning to analyze social data, from problem conception through model selection.  At each step, we identify critical issues that arise which can only be addressed with theory.

Throughout the section, two broad themes arise. First, given the breadth of data available to us, we sometimes act opportunistically; we use what we have to jump quickly on new problems.  This push to rapidly solve pressing social problems by using widely available data and methods leads us to, for example, use a dataset to answer a research question that the dataset is ill-suited for. Problems can arise from these decisions that entirely undermine the utility of the research---for example, selecting a bad sample of data can undermine the external validity of a study. 

Second, we often rely on intuition to make decisions about research design. For example, when constructing annotation tasks, intuition can lead to a overly simplified designs, when many other potential approaches could also be equally, or more, valid \cite{joseph_constance_2017-1}.  Often, these intuitions are good enough for the task at hand. However, when out intuitions are wrong, the results can be problematic. For example, following misguided intuitions about sexuality and its causes can lead to incorrect claims, made on top of poor research design decisions, about the biological nature of sexuality \cite{wang_deep_2018}.

This combination of opportunism and intuition can be particularly pernicious when combined with a lack of theory. While social scientists also often rely on intuition \cite{tavory2014abductive}, they can rely on the scaffolding provided by previous theoretical work, guiding them towards a better research design and/or understanding of their data. In Section~\ref{sec:theory_in}, we discuss how we can use social theory to help us constrain our opportunism and intuitions to existing knowledge of the social world provided to us by social theory.  This increases our chance at producing new, lasting science that helps move forward our understanding of society.

\subsection{Problem Selection and Framing}

As researchers, we are constantly asking ourselves, ``what problem should we be studying?''\footnote{Importantly, not \emph{all} researchers ask this question. Others, for example, may seek to understand the consequences of the practices they are studying. Thank you to stef shuster for pointing this out.} In core machine learning research, problems of interest revolve around technical innovation --- how can we develop new statistical or mathematical theory or approaches to improve the quality of our methods?  In the social sciences, we often focus instead on research that can help us gain new insights about the social world. 

Unfortunately, while technical approaches can sometimes help identify oddities in social data worth investigating, there is no technical solution to identifying good social science research questions.  These insights require an understanding of what is known already about the social world, and where gaps in this knowledge lie.  However, with the onslaught of big data, we all too often optimize for convenience, using the data we have on hand to solve problems just because they seem solvable, and because they seem to have real-world relevance. For example, we use publicly available Twitter data to predict people's movements within cities \cite{bauer_talking_2012} or aggregated search data from Google Trends to predict the prevalence of the flu \cite{lazer_parable_2014}. 

This convenience approach to problem selection and framing leads to two problems. First, it can lead us to formulate and tackle problems that seem important but in reality serve chiefly as an exercise in prediction, providing little new insight into the social world. Second, it can lead us to address problems that our intuitions accurately assume are important, but leave us struggling to frame the reasons \emph{why} the problem is important.  Social theory can help to alleviate these issues.

First, theory tells us which problems are worth solving. For example, election prediction is an essential research tool because it provides a model for understanding political processes \cite{beauchamp_predicting_2017,kennedy_improving_2017-1}. However, theory tells us that because of polarization, gerrymandering, and campaign finance laws, most American elections today are very predictable with only one piece of data – knowing who the incumbent is. Theory also tells us, however, that in nominally competitive races, \emph{polling} provides the next best predictor, because politics is driven by opinion. However, polling is expensive, and s only available for the most high-profile races. Theory thus suggests that within the domain of elections, the correct problem to study is modeling opinion in competitive and under-polled elections.

Second, theory can help us to motivate and frame problems that seem intuitively important. It may be apparent that predicting the prevalence of the flu can help save lives. However, less obvious is what problem is being solved when predicting, for example, a person's political affiliation based on their social media behavior (e.g. based on their tweets) \cite{cohen_classifying_2013}. Without theory, this prediction is simply an exercise in engineering. However, recent work on political polarization urges us to study affiliation as a function of partisan identity \cite{levendusky2009partisan}, and shows that such identities are rapidly undermining social and cultural stability in the United States \cite{doherty2017key}. Social theory therefore explains why predicting political affiliation is important – in order to study its association with cultural polarization \cite{dellaposta_why_2015}.

There are many examples where scholars using machine learning on social data have used theory to identify and frame important problems. For example, several scholars have addressed precisely the problem of opinion polling in competitive and under-polled elections using big data \cite{beauchamp_predicting_2017,wang_forecasting_2015}.  And Cranshaw et al. \cite{cranshaw_livehoods_2012} take an intuitively interesting task---clustering venues on foursquare by the patrons they have in common---and ground it in theory from urban studies and design to motivate and frame their work as addressing an unsolved problem of how neighborhood boundaries should be defined.

\subsection{Outcome Definition}

Having established a problem of interest, we turn to the task of defining and measuring our outcome. Our outcome measure is ideally meant to be the ground truth for our phenomenon we're modeling, i.e. an observation of the phenomenon itself. For example, if we are interested in studying partisanship, we can establish ground truth through a variety of means---whether someone votes for a single party \cite{poole1990patterns}, who they donate money to \cite{bonica_mapping_2014}, or what topics they tweet about \cite{tsur_frame_2015}. However, this data is often not easily available.  The easiest ``technical solution'' to this is simply to use a variable available in our data as ground truth, or, as we discuss in Section~\ref{sec:annotation}, to construct the variable through a rapid crowdsourced annotation task.   

However, no technical tool we are aware of can tell you whether or not, for example, a) voting behavior or b) political sentiment in tweets are valid or better measures of whether one is a liberal or conservative \cite{cohen_classifying_2013}. Answering this requires a theoretical understanding of what we mean by liberal and conservative. In turn, we must approach ground truth as something being theorized by researchers. Consequently, when we define an outcome, we have to do so using these theories to argue that the variable validly captures the construct we seek to measure \cite{hacking1986making}.  

With respect to liberalism and conservatism, for example, theory shows that people and sociotechnical systems shape the ``ground truth''. Only recently have liberal and conservative ideologies aligned with the Democratic and Republican parties in the United States – what is called partisan sorting \cite{mason_i_2015}.  For example, \cite{gentzkow2016measuring} show that partisan ideology has only become distinguishable in Congressional floor speeches since 1980. That is, language has only become partisan in the past forty years. 

These theoretical insights, in turn, helps us create valid outcomes by grounding its construction in a definition of what the outcome should be. Instead of predicting liberalism/conservatism, we likely would prefer to focus on if someone is a democrat or a republican. This is because partisan identity theory \cite{van_bavel_partisan_2018} claims that party membership is driven by party identification. What makes someone a democrat is not that they support public health care or market regulation but that they identify with democrats. Thus, if we want to infer someone’s political party identification from their tweets, we should look to who’s side they take in a debate, rather than the specific issues they support. In his campaign, Donald Trump famously supported liberal policies like public health care and criticized the war in Iraq. These stances did not make him a moderate conservative. They made him a populist republican as opposed to an establishment republican.

\subsection{Data Selection}

The process of data selection is defined as the identification of one or more datasets that can be used to address the problem under study.Data selection is typically carried out using either precedent or convenience as a heuristic.  Where possible, we seek out existing and established datasets that we can simply adopt for our own use. Where existing data do not exist, we often seek data that can be easily and scalably collected.  

This use of precedence and convenience stems from our interest not only in answering questions about the social world, but in desiring to do so via novel methodologies.  For example, when constructing novel solutions to (or models for) existing problems, we tend to reach for established datasets for which prior results exist for comparison. And for novel data collection, our methods often require large datasets, and so convenient collection of this data is almost a prerequisite.

But relying on either convenience or precedent can cause issues for social science questions, because all data contain both inclusions and exclusions that manifest in varying forms of bias \cite{olteanu_social_2016}.  By taking shortcuts with data selection, we often choose to ignore or brush over these inclusions and exclusions. For example, Blodgett et al. \cite{blodgett_demographic_2016} show that language identification tools shown to perform well on a wide array of text corpora \cite{lui2012langid} suffer significantly at distinguishing African-American English as English in social media texts. Because scholars have often used this tool to filter out non-English tweets, the result is a set of studies on social media data where the voices of African Americans are diminished relative to white Americans.

As Blodgett et al. \cite{blodgett_demographic_2016} suggest, socio-linguistic theory could have helped us to anticipate the potential issues in using the convenient language classifier they studied to identify English versus non-English content.  Specifically, theoretical models of how dialects form emphasize that variations of written English may not readily align in terms of the primary features used by the language classification model, n-grams of characters \cite{rickford1999african}.  Further, socio-linguistic theory emphasizing the importance of African American English and its distinctions from other English dialects in the presentation of the self online for Americans \cite{florini2014tweets} would have emphasized the need for social media scholars to reconsider the notion that there is a single definition of English that they wish to study.

More broadly, then, social theory helps us to understand the implications of how we make our decisions on what data to include or not include in our sample. This is especially critical when we expect others will reuse our data or the models we construct from them. For example, a  significant  amount  of  research  has  used  pre-trained  word  vectors  from  an unreleased Google  news  corpus, meaning the biases in the data are both unclear and unknown. On the contrary, Matt Salganik and colleagues \cite{lundberg2019privacy} use the statistical sampling and survey measurement theories baked into the Fragile Families Wellbeing Study to create the Fragile Families Challenge - a common data set computational social scientists can use to develop models predicting critical social factors like income, health, and housing. The use of theory to identify and explain important inclusion and exclusion variables have allowed research conducted during the challenge to contribute successfully to social scientific knowledge on child development \cite{salganik2019introduction}.

\subsection{Feature Engineering}

Feature engineering encompasses the process of converting our raw data to quantities that can be input into a machine learning model. The key question is, of course, how do we know that we have engineered the right features?  A technical solution to this question has typically privileged model performance. In supervised learning, we assess our feature engineering strategy by comparing predictive performance metrics (e.g. precision) across various potential feature sets. In unsupervised learning, we assess our strategy both by looking to other kinds of automatically-calculated or annotation-based performance metrics \cite{morstatter_search_2017}. 

The problem with this approach to feature engineering is that the features we select might boost our performance but may not help us distinguish genuine from spurious signal. Overfitting to noise is one way in which injudicious feature selection can inflate performance. Another is to include features that are spuriously related with our outcome of interest or exclude features that are directly related. Take the case of recidivism prediction as an example. To predict who will return to prison, not only do we need features that signal a person's propensity to commit a crime, but also features that capture police and judicial processes around who is likely to be arrested and convicted of a crime. Omitting features capturing both crime commission and arrest yields a poorly-specified model that performs well. 

Automated causal inference at scale is an as-yet unattained holy grail in machine learning. Thus, without theory, we cannot enumerate which features we should include and which we should exclude. Specifying the theoretical model in advance is the only way to enumerate what features we should generate \cite{pearl_seven_2019,rohrer_thinking_2018}. Building theoretical models allow us to identify which features should be included, and if they are deemed important by a model, what they might mean. More concretely,  Wallach \cite{wallach_computational_2018} argues that we should always be informing our selection of features and understanding of the problem with theory-based causal models in mind. This argument is, of course, at odds with claims of ``featureless'' models, as many claim deep learning models to be. For example, where before we may have needed to provide a model for Named Entity Recognition with part-of-speech tags for each input word, modern deep learning architectures do not require this feature engineering step \cite{goldberg_primer_2016}. However, even with such models, we are still making implicit decisions about our features, for example, by deciding whether to use words or characters as input to the model \cite{devlin2018bert}. Further, the causal processes of interest often lay beyond decisions on whether or not to use words or characters.  For example, regardless of what deep NLP model we choose to model an individual's language, word choices are often driven by more difficult-to-capture variables, like age and gender \cite{SchwartzPersonalityGenderAge2013a}.

\subsection{Annotation}\label{sec:annotation}

Oftentimes we cannot identify some critical feature we want to model from our data. For example, Twitter does not provide data on the gender or religious affiliation of their users. In such cases, we often ask humans, be it ourselves or others, to go through the data and identify the feature of interest by hand. This process of annotation has been understood as identifying some ground truth label that can then be used for model building. 

When annotating data, a primary goal is to ensure that annotators agree. Due to both noise and intrinsic variation amongst individuals, different people look at the same data and come up with different labels. Our interest, particularly when searching for some objective ground truth, is to ensure that despite these differences, we can identify some annotated value on which most annotators roughly agree. Scholars in the social sciences have long established statistical measures of agreement in annotation \cite{krippendorff2004reliability}, which are readily used in the machine learning pipeline. However, machine learning researchers have also sought to increase agreement in various ways \cite{snow_cheap_2008}.  These technical efforts to increase agreement largely rely on either trying to find the best annotators (i.e. those that tend to agree most frequently with others \cite{ipeirotis_repeated_2010}), finding better aggregation schemes \cite{passonneau_benefits_2014,raykar_learning_2010}, or simply by increasing the amount of data labeled \cite{snow_cheap_2008}.

At the core of many disagreements between annotators, however, is that the constructs we are seeking to annotate are often difficult to define.  For example, Joseph et al. \cite{joseph_exploring_2016} built a classifier to identify social identities in tweets, a concept that is notoriously varied in its meaning in the social sciences. Thus, even experts disagree on exactly what a social identity constitutes. Unsurprisingly, then, Joseph and his colleagues  found that non-expert annotators provided unreliable annotations, even after a discussion period. Annotations of hate speech have seen similar struggles, with limited agreement across annotators \cite{davidson2017automated} and with significant differences across annotators with different demographics \cite{waseem_are_2016}.

In such cases where the construct is difficult to define, technical solutions like adding more annotators or performing different aggregation schemes are unlikely to increase agreement. This is because, as with outcome definition,technical solutions cannot address the fundamental issue---defining the construct itself. In other words, technical solutions cannot be used to answer  the questions, ``what is a social identity?'' Or, ``what is hate speech?''  Instead, we must rely on theory to provide a definition. For example, Affect Control Theory in sociology focuses not on the general idea of social identity, but rather on ``cultural identity labels'', defined as ``(1) the role-identities indicating positions in the social structure, (2) the social identities indicating membership in groups, and (3) the category memberships that come from identification with some characteristic, trait, or attribute'' (\cite{smith-lovin_strength_2007-1}, pg. 110). Upon using this definition, and annotations from Affect Control theorists, Joseph et al. \cite{joseph_exploring_2016} noted a significant increase in annotation quality. 

Annotation, particularly with complex phenomena like identity, hate speech, or fake news \cite{grinberg_fake_2019}, therefore requires starting with a theory of the construct we wish to measure and its intersection with the subjective processes of our annotators.  One additional tool worth noting for this task that social scientists have developed is \emph{cognitive interviewing} \cite{beatty2007research}. Cognitive interviewing involves talking to potential annotators about how they think of the construct, its potential labels, how they would identify those labels, and then having them actually try to apply our task to some test data.  While similar to the idea of a pilot annotation task that machine learning researchers are likely familiar with, cognitive interviewing outlines specific ways in which theory can be applied before, during, and after the pilot to help shape the definition of the construct. Finally, although beyond the scope of the present work, it is also critical that annotation follows best methodological practices for structured content analysis in the social sciences \cite{geiger2019garbage}. 

\subsection{Model Construction}
In building a machine learning model for social data, our goal is to predict, describe, and/or explain some social phenomenon. Our job, then, is to identify the model that best accomplishes this goal, under some definition of best. Our challenge is to determine which of the many modeling approaches (e.g. a deep neural network versus a Random Forest) we can take, and which specific model(s) (e.g. which model architecture with what hyperparameters) within this broad array we will use for analysis.

It can be, and often is, overwhelming to select which model to use for a given analysis.  Consider, for example, the goal of understanding the topics in a corpora of text. The early topic modeling work of Blei et al. \cite{blei_latent_2003}, has been cited over 28,000 times. Many of these citations are from extensions of the original model. For example, there are topic models for incorporating author characteristics \cite{rosen2004author}, author characteristics and sentiment \cite{MukherjeeJointAuthorSentiment2014}, author community \cite{liu_topic-link_2009}, that deal specifically with short text \cite{yan2013biterm}, that incorporate neural embeddings of words \cite{card_neural_2017}, and that emphasize sparsity \cite{EisensteinSparseadditivegenerative2011a}.  How do we construct a model that is best, or right, for our analysis?

Brendan O'Connor, David Bamman, and Noah Smith \cite{oconnor_computational_2011} describe this kind of modeling choice as occurring along two axes - computational complexity and domain assumptions.  Computational complexity is used loosely to represent complexity in computational time and  ``horsepower''. Domain assumptions vary from few assumptions, essentially assuming ``the model will learn everything,'' to cases where we explicitly model theory. However, O’Connor et al. leave open the question of where in this space the ``right'' model for a particular problem is likely to fall, or how to define the right domain assumptions.

This is where theory comes in. By defining the goal of the model – prediction, explanation, description, and so on; and providing clear expectations for what our domain assumptions are, theory helps us navigate the computation/domain space.  In the context of topic modeling, the Structural Topic Model (STM) \cite{roberts_structural_2013-1,roberts_structural_2014} provides a generic framework for defining our domain assumptions based on the factors we expect to be important for shaping the topics that appear in a document. By incorporating covariates into the modeling process that we theorize to be relevant, we can leverage theory both to create a model that ``fits the data better,'' and get outputs of the model that we can use to directly test extensions to our theory. The right model, then, is defined by theory. For example, Farrell \cite{farrell2016corporate} uses theories of polarization through ``contrarian campaigns'' that originate in well-funded organizations to determine a particular instantiation of the Structural Topic Model that they use to study how polarization has emerged on the topic of climate change.

The STM is therefore useful in that, given an established set of generic modeling assumptions and a defined level of computational complexity, we can use theory to define the specific model we construct.  Similar efforts have been made in other areas of text analysis as well. For example, Hovey and Fornaciari \cite{hovy_increasing_2018} use the concept of homophily, that people with similar social statuses use similar language, to retrofit their word embedding model. This theory-driven change allowed the model to leverage new information, yielding a more performant model.  As such, the use of theory to guide natural language processing models can serve as a blueprint for the application of theory in other domains of social data.

\section{Theory Out} 

Machine learning has traditionally concerned itself with maximizing predictive performance. This means that the first results reported in machine learning papers, those in ``Table 1'', are often a report on the model's predictive performance relative to some baselines.  However, scholars are increasingly interested in other aspects of model output, like interpretability and fairness. In applied research, it is important for scholars to demonstrate that their model helps us understand the data and explains why particular predictions are made.  These new demands for the output of machine learning models create problems for which technical solutions have been proposed.  In this section, we argue that this technical innovation must be supplemented with social theory if we are to truly interpret, explain, and use our models to learn about and improve the social world. 

\subsection{Interpretability, Explainability, and  Theory-building}

Few criticisms have been leveled against machine learning models more than the charge that they are uninterpretable.  While a concrete definition of interpretability has been elusive \cite{lipton_mythos_2016}, the general critique has been that machine learning models are often``black boxes'', performing complex and unobservable procedures that produce outputs we are expected to trust and use. In trying to open the black box and account for our models, three distinct questions are often treated interchangeably:

\begin{itemize}
\item \emph{ {\bf What} did the model learn, and how well did it learn it?}  Meaning, given a particular input, how does the model translate this to an output and how accurately does this output match what we expect? We refer to this as the question of \emph{interpretability}.
\item \emph{ {\bf Why} did the model learn this?} What is it about the (social) world that led to the model learning these particular relationships between inputs and outputs?  We will refer to this as the question of \emph{explainability}.
\item \emph{What did we learn {\bf about the world} from this model?} What new knowledge about the social world can be gleaned from the results of our model? We refer to this as the question of \emph{theory-building}.
\end{itemize}

Interpretability, explainability, and theory-building get lumped together in the technical solutions that have been developed to open the black box. For example, sparsity-inducing mechanisms like regularization \cite{friedman2009glmnet} and attention (in neural networks; \cite{vaswani2017attention}) increase interpretability by minimizing the number of parameters to inspect. In turn, these technical solutions are used help us explain how the parameters relate to the data generating process (explainability) \cite{zagoruyko_paying_2016}. Similarly, qualitative analysis of what the model got right and wrong through case studies and error analysis sheds light on both the limits of what the model has learned (interpretability) and why it makes incorrect predictions (explainability). We also use model-based simulations, tweaking inputs to show how they produce different outputs \cite{ribeiro_why_2016} and adversarial examples that fool the model to explore its performance (interpretability) the limits of its understanding about the world (theory-building) \cite{wallace_universal_2019}. 

However, while there are many methodological overlaps; interpretation, explanation, and theory-building are distinct research questions that require distinct practices with distinct roles for social theory.

When interpreting models-- that is, when trying to understand what the model learned and how well it learned it--- social theory enables us to go beyond the technical question of \emph{how} to look at the model to \emph{what to look at}. In order to choose what model parameters, characteristics, and cases to visualize, we need to have theory-driven expectations about how the model is \emph{supposed} to work and what it is \emph{supposed} to do. For example, social theory tells us that racial bias is tied to visual cues beyond just skin color. Bias driven by skin tone, called `colorism,’ is different from bias driven by cultural codes around race. For example, the color of someone's skin biases people in ways that are different from how they are dressed, their accent, and hair style \cite{todorov2008understanding}. Consequently, if we want to interpret the biases of a computer vision model through the lens of color and race, social theory tells us we should look to see whether our machine vision models are more sensitive to skin tone or cultural indicators.  This can help differentiate, for example, whether biases in the model derive from cultural norms embedded in the population that make up the training data or from under-representation of individuals with certain skin tones in the training data (or both) \cite{hanna_towards_2019,benthall_racial_2019-1}.

A good example of how theory can be used to guide interpretation is the work from Bamman et al. \cite{bamman_bayesian_2014}, who identify tropes of literary characters. They validate their model by testing specific parameters against a slate of theory-based hypotheses. These hypotheses, derived from theory on the writing styles of authors during the time period of study, took the form of “character X is more similar to character Y than either X or Y is to a distractor character Z.”  Good models were those that accurately predicted these theorized character similarities. By combining visualization, summarization, and other interpretive techniques to test different theories, we develop an account of the world that is consistent with established research rather than what is obvious \cite{watts_everything_2011}.

Oftentimes, we take our interpretation of model behavior and develop an account of why the model did what it did based on that interpretation.  This often serves as an explanation of what the model did and an effort to build new theory.  However, when we build explanations and new theory only on insights gleaned from examining model behavior, we are at risk of developing \emph{folk theory} \cite{d1995development}. Folk theory involves leaning on common understanding to ``read the tea leaves'' \cite{ChangReadingTeaLeaves2009} characterizing human behavior as simply ``making sense.'' This is dangerous, however \cite{kerr_harking_1998}. Models will always output \emph{something} and some model will always outperform others on some metrics. Building theory only from numerical outputs serves only to reinforce common myths and biases. 

If we would like for our explanations of our model to contribute to a broader understanding of the social world, we need to not only find the right explanation for each model, but to also integrate many models and explanations into a coherent account of the world. Nelson’s work on the development of second wave feminism is a prime example. She used social network and feminist theory to build different machine-learning based models for the structure of feminist communities in New York and Chicago. She then compared the structures of the social and idea networks to show that the ideas central to feminist community in New York were more aligned with what we understand today to be ``second wave’’ feminism and that their community was more densely connected than that in Chicago. She argues this dense connectivity enabled feminists in New York to set the agenda for feminism in the 1960s and 70s. 

\subsection{Generalizability}
 
Generalizability refers to the goal of understanding how well results apply to the cases that were not tested. For example, if we develop a model to predict unemployment using mobile phone data in Europe \cite{toole_jameson_l._tracking_2015}, an analysis of generalizability might involve assessing whether the same approach would work in Algeria, Canada, or Mexico. We might also like to know how likely it is to work on mobile phone data from five years ago or five years in the future. And, we might want to know whether a similar kind of approach would work on other data sets like social media posts, internet searches, or transit data.

In machine learning, generalizability is often addressed technically by reapplying the same methodology to other data to see whether it performs similarly to the original.  For example, generalizability of a particular modeling approach like topic models might be tested by applying a fitted model to different kinds of data and different problems. We also test the generalizability of a particular analytic approach by reapplying it in different domains. For example, Lucas et al. \cite{lucas_computer-assisted_2015} use machine translation across multiple languages to study whether politics in different countries were constituted by the same issues being discussed in the same ways. 

Beyond mechanical tests, machine learning researchers have developed automated procedures to try to encourage generalizability of their models. Many of these approaches are ways to mitigate overfitting, which can improve model generalizability by reducing the extent to which a model learns from patterns unique to the given data and/or feature set \cite{hastie2009elements}. Recently, efforts have been made to train a model that learns representations of some universal input which can then be fine-tuned to apply to a variety of problems. For example, ResNet \cite{szegedy2017inception} and BERT \cite{devlin2018bert} learn generic representations for images and sentences, respectively, and can then be fine-tuned for various classification tasks. 

However, while these technical solutions can make individual models more generalizable, they cannot help us establish why a result on one dataset can be generalized (or not) to others. For this, we need theory to tell us what similarities and differences are salient. Zeynep Tufekci \cite{tufekci_big_2014-1} makes this point when arguing that we cannot treat one online platform (i.e. Twitter) as a stand-in for all others – as a \emph{model organism} for society. The platform rules, social dynamics, and population that make Twitter worth engaging in for its users also distinguish it fundamentally from services like Facebook, Instagram, and WhatsApp. For example, theories of homophily suggests that, on any platform, people will associate with others like them. Yet, the commonalities on which we build connections depend on the platform itself. Our friends, colleagues, and public figures are on Twitter and our family is on Facebook. Following Goffman's theory of presentation of self, these differences in audiences drive people to behave differently on different platforms \cite{goffman_presentation_1959}.

Moving forward then, we see theory as a way to develop paradigms for understanding for particular kinds of data, like data from different social media platforms, and for how models might be tweaked to generalize beyond the data they were trained on. 

\subsection{Parsimony}

Parsimony refers to the goal of building a model with as few parameters as possible while still maximizing performance. Machine learning models benefit from parsimony because it decreases complexity and cost, limits the danger of overfitting, and makes it easier to visualize \cite{hastie2009elements}.

A variety of technical approaches for constructing parsimonious models exist. For example, we can use topics, clustering, or factoring to reduce feature dimensionality, and regularization to remove weak predictors. In the case of neural networks, we also use techniques like drop-out \cite{gal2016dropout} or batch normalization \cite{ioffe2015batch}.

There are, however, three flaws with these approaches to addressing parsimony. First, because many features are correlated with one another and the outcome, technical approaches often arbitrarily select certain correlated features and not others. This arbitrary selection can make it difficult to differentiate between truly irrelevant features and those that are simply correlated with other relevant features. Second, decisions on when the model is ``parsimonious enough'' rely largely on heuristic comparisons between model performance on training and validation data (e.g. the ``1-Standard Error rule'' used in the popular \texttt{glmnet} package in \texttt{R} \cite{friedman2009glmnet}). Such approaches cannot help guide our intuitions on what number of predictors ``makes sense''. Finally, the standard machine learning assumption that we need many features can be incorrect even at relatively low values. It is often the case in social science problems that a small set of variables can easily explain a large part of the variance. A regularizer may select 1,000 features out of 10,000 while the best model may only need 50.

Social theory provides a solution to these issues by helping us define small sets, or ``buckets,'' of variables that we expect to explain a large portion of the variance in the outcome. Instead of starting with many features and trying to weed out irrelevant ones, theory points us in the direction of the most important variables, helping establish a baseline level of predictability from which we can assess whether additional features provide additional performance. Similarly, because theory provides us with the features we expect to be important, we may be able to identify cases in which regularization removes important, stable predictors due to correlation amongst variables. 

The idea of identifying parsimonious, theoretically-informed baseline models for comparison has been shown to work well in practice. In their study of Twitter cascades, Goel et al. \cite{goel_structural_2015} show that a simple model which accounts only for popularity of the user is an extremely strong baseline for predicting the size of a retweet cascade. These ideas align with theories of source credibility \cite{hovland1951influence} and the importance of influential nodes in information spread \cite{marsden1993network} that emphasize the centrality of social network structure in information flow. Similarly, social homophily has also long provided simple yet powerful baselines for recommendation systems \cite{mcpherson_birds_2001}. 

In linguistics, simple theory-driven models of have routinely been shown to have high levels of predictive validity. For example, as Bender notes \cite{bender2016linguistic}, 
theories of linguistic variation across languages are critical for the expansion of existing English-only techniques to a broader array of languages for which the same level of training data and/or knowledge is not available. Combined, these efforts have informed the development of more formal social theory on the limits of predictability in social systems \cite{hofman_prediction_2017}, which may further extend our ability to estimate the degree of parsimony expected for particular problems.

\subsection{Fairness}

In both popular media \cite{li_aoc_2019} and academic literature \cite{mitchell_prediction-based_2018}, significant attention has turned to the question of how machine learning models may lead to increased discrimination against certain social groups. This has translated into two areas of work.  First, scholars have focused on the construction of \emph{fair} machine learning models which purport to alleviate discrimination in model outputs altogether.  Second, a significant amount of research has focused on measuring fair\emph{ness} of the output. In both, fairness is defined at either the group level, such that there is no difference in outcome between groups, or at the individual level, such that individuals are treated similarly regardless of group status \cite{hutchinson_50_2019}.

A host of technical approaches have been developed to construct fair models and measure of fairness. For a recent overview on the relationship of this research to decades-old discussions of fairness in statistical models, see \cite{hutchinson_50_2019} or \cite{mitchell_prediction-based_2018}.  The bulk of the work to ensure fair outcomes has focused on batch classification and various manipulations of either data input to the model (e.g. de-biasing data according to groups \cite{kearns_preventing_2017}) or modifications to existing models (e.g. via a regularization term ensuring individual fairness \cite{kamishima2011fairness}) to ensure fair outcomes. On the metric construction side, scholars have recently focused on developing measures that account for sociologically relevant phenomena like intersectionality\footnote{Although see the critique from Hoffmann of this work \cite{hoffmann_where_2019}} \cite{foulds_intersectional_2018}, on the tradeoffs between existing measures \cite{kleinberg2018inherent}, and on a better understanding of the causal assumptions of different measures \cite{glymour_measuring_2019} amongst other tasks.

However, an increasing number of scholars have identified important complications to defining fairness technically. This work has emphasized 1) that different people have different views on what is fair, 2) that the views of those in power are the views that are most likely to be used, and 3) that models emerge from a vast and complex sociotechnical landscape where discrimination emerges from many other places beyond the models themselves \cite{crawford_can_2016,selbst_fairness_2018,barocas_social_2017,green_fair_2018,hoffmann_where_2019}. One conclusion has been that a fair algorithm cannot fix a discriminatory process. For example, recidivism prediction algorithms will almost certainly be used in a discriminatory fashion, regardless of whether or not the models themselves are fair \cite{green_fair_2018}. We need social theory to better understand the social processes in which these algorithms are embedded. Social theory enables us to distinguish discrimination caused by the algorithm from that originating in the social system itself. 

Perhaps equally important, theory can also help us to understand the \emph{consequences} of unfair and/or biased algorithms. For example, take recent work showing that search algorithms return gender and race stereotypical images for various occupations \cite{kay2015unequal}. Social psychology theories focusing on representation emphasize that from a young age, we internalize representations of occupations and skills that cause us to shift towards those that are stereotypical of our own perceived gender \cite{bian_gender_2017}. Thus, while technical solutions may help us to identify such problems, they cannot explain the impacts of these biases and thus why they should be addressed and how. 

Finally, social theory helps to identify how unfair machine learning impacts our knowledge about the world. Biased algorithms, such as those that detect gender and race for demographic comparisons \cite{jung2017inferring}, can bias the science we produce. Standpoint theory and other critical epistemological theories have shown how who does science and who's data are used for what analysis affects what we know about the social world \cite{harding2004feminist,haraway1988situated}. We do not want to replicate the patterns of exclusion and stigmatization found in the history of medicine \cite{martin_egg_1991}, psychology \cite{foucault1990history}, and sociology \cite{zuberi2008white} by throwing out data from marginalized people, only studying marginalized people as the Other, or not allowing marginalized people speak for themselves about their data.

\section{Conclusion}

The combination of machine learning methods and big social data offers us an exciting array of scientific possibilities. However, work in this area too often privileges machine learning models that perform well over models that are founded in a deeper understanding of the society under study. At best, this trade-off puts us in danger of advancing only computer science rather than both computer science and social science.  At worst, these efforts push the use of machine learning for social data towards pseudoscience, where misappropriated algorithms are deployed to make discriminatory decisions and baseless social scientific claims are made.

However, as the many positive examples we have highlighted here show, machine learning and big social data can be used to produce important, ground-breaking research. To do so, the examples we highlight have baked social theory into each step of the machine learning pipeline.  These works do not cherry-pick one theory, ex post facto, to support their claims. Instead, they use multiple, potentially competing theor\emph{ies}, at every step of the pipeline, to justify their inputs and help validate their outputs. In using, or at least acknowledging, competing theories, we can elucidate where disagreements exist and therefore which technical trade-offs are most important.

The positive examples we highlight, our review of the negative cases, and the related work we draw on pave the way forward for the scientifically-grounded, ethical application of machine learning to social data. But our efforts must move beyond the way we produce research to the ways we review it, consume it, and encourage it as a research community.  As reviewers, for example, we must ask ourselves if the work we are looking at is justified not only by statistical theory, but by social theory as well. And as a community, we must find ways to feature and promote papers that may not have the flashiest ``Table 1,'' but that provide a careful and well-grounded social scientific study.

Machine learning can and should become a critical piece of social science. The solution does not necessarily require a computer scientist to ``go find a social scientist,'' or vice versa. There is already a wealth of knowledge to draw from, and we should not allow ourselves or others to avoid delving into it simply because it is ``out of our field.''  For those who do not know where to start, we hope this paper is a guide to anyone for how to use that knowledge to address specific questions in the research. Similarly, social science should become an increasingly important part of machine learning. In incorporating social theory into their work, machine learning researchers need not reliquish model performance as the ultimate goal; we have argued here that, instead, theory can help guide the path to even better models and predictive performance.

\section{Acknowledgements}

We greatly appreciate the assistance of Laura Nelson, Celeste Campos-Casillo, stef shuster, Atri Rudra, and David Lazer, who all provided invaluable feedback on earlier versions of this work. That said, these individuals of course bear no responsibility for the current content; all issues, errors, and omissions are the fault of the authors alone.

\bibliographystyle{plain}
\bibliography{references}

\begin{thebibliography}{100}

\bibitem{abbott_sequence_1995}
Andrew Abbott.
\newblock Sequence {{Analysis}}: {{New Methods}} for {{Old Ideas}}.
\newblock {\em Annual Review of Sociology}, 21(1):93--113, August 1995.

\bibitem{abbott_transcending_1988}
Andrew Abbott.
\newblock Transcending {{General Linear Reality}}.
\newblock {\em Sociological Theory}, 6(2):169, 23.

\bibitem{arcas_physiognomys_2017}
Blaise Aguera~y Arcas, Margaret Mitchell, and Alexander Todorov.
\newblock Physiognomy's {{New Clothes}}, 2017.

\bibitem{bamman_bayesian_2014}
David Bamman, Ted Underwood, and Noah~A. Smith.
\newblock A bayesian mixed effects model of literary character.
\newblock In {\em Proceedings of the 52st {{Annual Meeting}} of the
  {{Association}} for {{Computational Linguistics}} ({{ACL}}'14)}, 2014.

\bibitem{barocas_social_2017}
Solon Barocas, danah {boyd}, Sorelle Friedler, and Hanna Wallach.
\newblock Social and {{Technical Trade}}-{{Offs}} in {{Data Science}}.
\newblock {\em Big Data}, 5(2):71--72, June 2017.

\bibitem{bauer_talking_2012}
S.~Bauer, A.~Noulas, D.O. Seaghdha, S.~Clark, and C.~Mascolo.
\newblock Talking {{Places}}: {{Modelling}} and {{Analysing Linguistic
  Content}} in {{Foursquare}}.
\newblock In {\em Privacy, {{Security}}, {{Risk}} and {{Trust}} ({{PASSAT}}),
  2012 {{International Conference}} on and 2012 {{International Confernece}} on
  {{Social Computing}} ({{SocialCom}})}, pages 348--357, 2012.

\bibitem{beatty2007research}
Paul~C Beatty and Gordon~B Willis.
\newblock Research synthesis: The practice of cognitive interviewing.
\newblock {\em Public opinion quarterly}, 71(2):287--311, 2007.

\bibitem{beauchamp_predicting_2017}
Nicholas Beauchamp.
\newblock Predicting and {{Interpolating State}}-{{Level Polls Using Twitter
  Textual Data}}.
\newblock {\em American Journal of Political Science}, 61(2):490--503, 2017.

\bibitem{bender2016linguistic}
Emily~M Bender.
\newblock Linguistic typology in natural language processing.
\newblock {\em Linguistic Typology}, 20(3):645--660, 2016.

\bibitem{benthall_racial_2019-1}
Sebastian Benthall and Bruce~D. Haynes.
\newblock Racial categories in machine learning.
\newblock In {\em Proceedings of the {{Conference}} on {{Fairness}},
  {{Accountability}}, and {{Transparency}}}, pages 289--298. {ACM}, 2019.

\bibitem{bian_gender_2017}
Lin Bian, Sarah-Jane Leslie, and Andrei Cimpian.
\newblock Gender stereotypes about intellectual ability emerge early and
  influence children's interests.
\newblock {\em Science}, 355(6323):389--391, 2017.

\bibitem{blei_latent_2003}
David~M. Blei, Andrew~Y. Ng, and Michael~I. Jordan.
\newblock Latent dirichlet allocation.
\newblock {\em J. Mach. Learn. Res.}, 3:993--1022, March 2003.

\bibitem{blodgett_demographic_2016}
Su~Lin Blodgett, Lisa Green, and Brendan O'Connor.
\newblock Demographic dialectal variation in social media: {{A}} case study of
  {{African}}-{{American English}}.
\newblock {\em EMNLP'16}, 2016.

\bibitem{bolukbasi_man_2016-1}
Tolga Bolukbasi, Kai-Wei Chang, James~Y. Zou, Venkatesh Saligrama, and Adam~T.
  Kalai.
\newblock Man is to computer programmer as woman is to homemaker? debiasing
  word embeddings.
\newblock In {\em Advances in {{Neural Information Processing Systems}}}, pages
  4349--4357, 2016.

\bibitem{bonica_mapping_2014}
Adam Bonica.
\newblock Mapping the {{Ideological Marketplace}}.
\newblock {\em American Journal of Political Science}, 58(2):367--386, 2014.

\bibitem{card_neural_2017}
Dallas Card, Chenhao Tan, and Noah~A. Smith.
\newblock A {{Neural Framework}} for {{Generalized Topic Models}}.
\newblock {\em arXiv preprint arXiv:1705.09296}, 2017.

\bibitem{ChangReadingTeaLeaves2009}
Jonathan Chang, Jordan~L. {Boyd-Graber}, Sean Gerrish, Chong Wang, and David~M.
  Blei.
\newblock Reading {{Tea Leaves}}: {{How Humans Interpret Topic Models}}.
\newblock In {\em {{NIPS}}}, volume~22, pages 288--296, 2009.

\bibitem{cohen_classifying_2013}
Raviv Cohen and Derek Ruths.
\newblock Classifying political orientation on {{Twitter}}: {{It}}'s not easy!
\newblock In {\em {{ICWSM}}}, 2013.

\bibitem{cranshaw_livehoods_2012}
Justin Cranshaw, Raz Schwartz, Jason~I. Hong, and Norman Sadeh.
\newblock The {{Livehoods Project}}: {{Utilizing Social Media}} to
  {{Understand}} the {{Dynamics}} of a {{City}}.
\newblock In {\em Proceedings of the {{Sixth International AAAI Conference}} on
  {{Weblogs}} and {{Social Media}}}, {{ICWSM}} '12. {AAAI}, 2012.

\bibitem{crawford_can_2016}
Kate Crawford.
\newblock Can an {{Algorithm}} be {{Agonistic}}? {{Ten Scenes}} from {{Life}}
  in {{Calculated Publics}}.
\newblock {\em Science, Technology, \& Human Values}, 41(1):77--92, January
  2016.

\bibitem{crawford_ai_2019}
Kate Crawford, Roel Dobbe, Dryer Theodora, Genevieve Fried, Ben Green,
  Elizabeth Kaziunas, Amba Kak, Varoon Mathur, Erin McElroy, Andrea
  Nill~Sanchez, Deborah Raji, Joy~Lisi Rankin, Rashida Richardson, Jason
  Schultz, Sarah Myers~West, and Meredith Whittaker.
\newblock {{AI Now}} 2019 {{Report}}.
\newblock 2019.

\bibitem{d1995development}
Roy~G d'Andrade.
\newblock {\em The development of cognitive anthropology}.
\newblock Cambridge University Press, 1995.

\bibitem{davidson2017automated}
Thomas Davidson, Dana Warmsley, Michael Macy, and Ingmar Weber.
\newblock Automated hate speech detection and the problem of offensive
  language.
\newblock In {\em Eleventh international aaai conference on web and social
  media}, 2017.

\bibitem{dellaposta_why_2015}
Daniel DellaPosta, Yongren Shi, and Michael Macy.
\newblock Why {{Do Liberals Drink Lattes}}?
\newblock {\em American Journal of Sociology}, 120(5):1473--1511, March 2015.

\bibitem{devlin2018bert}
Jacob Devlin, Ming-Wei Chang, Kenton Lee, and Kristina Toutanova.
\newblock Bert: Pre-training of deep bidirectional transformers for language
  understanding.
\newblock {\em arXiv preprint arXiv:1810.04805}, 2018.

\bibitem{doherty2017key}
Carroll Doherty.
\newblock Key takeaways on americans’ growing partisan divide over political
  values.
\newblock {\em Pew Research Center}, 2017.

\bibitem{EisensteinSparseadditivegenerative2011a}
Jacob Eisenstein, Amr Ahmed, and Eric~P. Xing.
\newblock Sparse additive generative models of text.
\newblock In {\em Proceedings of the 28th {{International Conference}} on
  {{Machine Learning}} ({{ICML}}-11)}, pages 1041--1048, 2011.

\bibitem{farrell2016corporate}
Justin Farrell.
\newblock Corporate funding and ideological polarization about climate change.
\newblock {\em Proceedings of the National Academy of Sciences}, 113(1):92--97,
  2016.

\bibitem{florini2014tweets}
Sarah Florini.
\newblock Tweets, tweeps, and signifyin’ communication and cultural
  performance on “black twitter”.
\newblock {\em Television \& New Media}, 15(3):223--237, 2014.

\bibitem{foucault1990history}
Michel Foucault.
\newblock {\em The history of sexuality: An introduction}.
\newblock Vintage, 1990.

\bibitem{foulds_intersectional_2018}
James Foulds and Shimei Pan.
\newblock An {{Intersectional Definition}} of {{Fairness}}.
\newblock {\em arXiv preprint arXiv:1807.08362}, 2018.

\bibitem{friedman2009glmnet}
Jerome Friedman, Trevor Hastie, and Rob Tibshirani.
\newblock glmnet: Lasso and elastic-net regularized generalized linear models.
\newblock {\em R package version}, 1(4), 2009.

\bibitem{gal2016dropout}
Yarin Gal and Zoubin Ghahramani.
\newblock Dropout as a bayesian approximation: Representing model uncertainty
  in deep learning.
\newblock In {\em international conference on machine learning}, pages
  1050--1059, 2016.

\bibitem{geiger2019garbage}
R~Stuart Geiger, Kevin Yu, Yanlai Yang, Mindy Dai, Jie Qiu, Rebekah Tang, and
  Jenny Huang.
\newblock Garbage in, garbage out? do machine learning application papers in
  social computing report where human-labeled training data comes from?
\newblock {\em arXiv preprint arXiv:1912.08320}, 2019.

\bibitem{gentzkow2016measuring}
Matthew Gentzkow, Jesse Shapiro, Matt Taddy, et~al.
\newblock Measuring polarization in high-dimensional data: Method and
  application to congressional speech.
\newblock Technical report, 2016.

\bibitem{glymour_measuring_2019}
Bruce Glymour and Jonathan Herington.
\newblock Measuring the {{Biases That Matter}}: {{The Ethical}} and {{Casual
  Foundations}} for {{Measures}} of {{Fairness}} in {{Algorithms}}.
\newblock In {\em Proceedings of the {{Conference}} on {{Fairness}},
  {{Accountability}}, and {{Transparency}}}, {{FAT}}* '19, pages 269--278, {New
  York, NY, USA}, 2019. {ACM}.

\bibitem{goel_structural_2015}
Sharad Goel, Ashton Anderson, Jake Hofman, and Duncan~J. Watts.
\newblock The {{Structural Virality}} of {{Online Diffusion}}.
\newblock {\em Management Science}, 62(1):180--196, July 2015.

\bibitem{goffman_presentation_1959}
Erving Goffman.
\newblock {\em The {{Presentation}} of {{Self}} in {{Everyday Life}}}.
\newblock {Anchor}, June 1959.

\bibitem{goldberg_primer_2016}
Yoav Goldberg.
\newblock A primer on neural network models for natural language processing.
\newblock {\em Journal of Artificial Intelligence Research}, 57:345--420, 2016.

\bibitem{gould1996mismeasure}
Stephen~Jay Gould, Steven~James Gold, et~al.
\newblock {\em The mismeasure of man}.
\newblock WW Norton \& Company, 1996.

\bibitem{green_fair_2018}
Ben Green.
\newblock ``{{Fair}}'' {{Risk Assessments}}: {{A Precarious Approach}} for
  {{Criminal Justice Reform}}.
\newblock In {\em 5th {{Workshop}} on {{Fairness}}, {{Accountability}}, and
  {{Transparency}} in {{Machine Learning}}}, 2018.

\bibitem{grinberg_fake_2019}
Nir Grinberg, Kenneth Joseph, Lisa Friedland, Briony {Swire-Thompson}, and
  David Lazer.
\newblock Fake news on {{Twitter}} during the 2016 {{U}}.{{S}}. presidential
  election.
\newblock {\em Science}, 363(6425):374--378, January 2019.

\bibitem{hacking1986making}
Ian Hacking.
\newblock Making up people.
\newblock 1986.

\bibitem{hanna_towards_2019}
Alex Hanna, Emily Denton, Andrew Smart, and Jamila {Smith-Loud}.
\newblock Towards a {{Critical Race Methodology}} in {{Algorithmic Fairness}}.
\newblock {\em arXiv preprint arXiv:1912.03593}, 2019.

\bibitem{haraway1988situated}
Donna Haraway.
\newblock Situated knowledges: The science question in feminism and the
  privilege of partial perspective.
\newblock {\em Feminist studies}, 14(3):575--599, 1988.

\bibitem{harding2004feminist}
Sandra~G Harding.
\newblock {\em The feminist standpoint theory reader: Intellectual and
  political controversies}.
\newblock Psychology Press, 2004.

\bibitem{hastie2009elements}
Trevor Hastie, Robert Tibshirani, and Jerome Friedman.
\newblock {\em The elements of statistical learning: data mining, inference,
  and prediction}.
\newblock Springer Science \& Business Media, 2009.

\bibitem{hipp_measuring_2012}
John~R. Hipp, Robert~W. Faris, and Adam Boessen.
\newblock Measuring `neighborhood': {{Constructing}} network neighborhoods.
\newblock {\em Social Networks}, 34(1):128 -- 140, 2012.

\bibitem{hoffmann_where_2019}
Anna~Lauren Hoffmann.
\newblock Where fairness fails: {{On}} data, algorithms, and the limits of
  antidiscrimination discourse.
\newblock {\em Information, Communication, and Society}, 2019.

\bibitem{hofman_prediction_2017}
Jake~M. Hofman, Amit Sharma, and Duncan~J. Watts.
\newblock Prediction and explanation in social systems.
\newblock {\em Science}, 355(6324):486--488, February 2017.

\bibitem{hovland1951influence}
Carl~I Hovland and Walter Weiss.
\newblock The influence of source credibility on communication effectiveness.
\newblock {\em Public opinion quarterly}, 15(4):635--650, 1951.

\bibitem{hovy_increasing_2018}
Dirk Hovy and Tommaso Fornaciari.
\newblock Increasing {{In}}-{{Class Similarity}} by {{Retrofitting Embeddings}}
  with {{Demographic Information}}.
\newblock In {\em Proceedings of the 2018 {{Conference}} on {{Empirical
  Methods}} in {{Natural Language Processing}}}, pages 671--677, {Brussels,
  Belgium}, October 2018. {Association for Computational Linguistics}.

\bibitem{hutchinson_50_2019}
Ben Hutchinson and Margaret Mitchell.
\newblock 50 {{Years}} of {{Test}} ({{Un}})fairness: {{Lessons}} for {{Machine
  Learning}}.
\newblock {\em Proceedings of the Conference on Fairness, Accountability, and
  Transparency - FAT* '19}, pages 49--58, 2019.

\bibitem{ioffe2015batch}
Sergey Ioffe and Christian Szegedy.
\newblock Batch normalization: Accelerating deep network training by reducing
  internal covariate shift.
\newblock {\em arXiv preprint arXiv:1502.03167}, 2015.

\bibitem{ipeirotis_repeated_2010}
Panagiotis~G. Ipeirotis, Foster Provost, Victor Sheng, and Jing Wang.
\newblock Repeated {{Labeling Using Multiple Noisy Labelers}}.
\newblock {\em SSRN eLibrary}, September 2010.

\bibitem{jacobs2019measurement}
Abigail~Z. Jacobs and Hanna Wallach.
\newblock Measurement and fairness, 2019.

\bibitem{joseph_constance_2017-1}
Kenneth Joseph, Lisa Friedland, William Hobbs, David Lazer, and Oren Tsur.
\newblock {{ConStance}}: {{Modeling Annotation Contexts}} to {{Improve Stance
  Classification}}.
\newblock In {\em Proceedings of the 2017 {{Conference}} on {{Empirical
  Methods}} in {{Natural Language Processing}}}, pages 1115--1124, Copenhagen,
  Denmark, September 2017. {Association for Computational Linguistics}.

\bibitem{joseph_exploring_2016}
Kenneth Joseph, Wei Wei, and Kathleen~M. Carley.
\newblock Exploring patterns of identity usage in tweets: A new problem,
  solution and case study.
\newblock In {\em Proceedings of the 25th {{International Conference}} on
  {{World Wide Web}}}, pages 401--412. {International World Wide Web
  Conferences Steering Committee}, 2016.

\bibitem{jung2017inferring}
Soon-Gyo Jung, Jisun An, Haewoon Kwak, Joni Salminen, and Bernard~J Jansen.
\newblock Inferring social media users demographics from profile pictures: A
  face++ analysis on twitter users.
\newblock In {\em Proceedings of 17th International Conference on Electronic
  Business}, 2017.

\bibitem{kamishima2011fairness}
Toshihiro Kamishima, Shotaro Akaho, and Jun Sakuma.
\newblock Fairness-aware learning through regularization approach.
\newblock In {\em 2011 IEEE 11th International Conference on Data Mining
  Workshops}, pages 643--650. IEEE, 2011.

\bibitem{kay2015unequal}
Matthew Kay, Cynthia Matuszek, and Sean~A Munson.
\newblock Unequal representation and gender stereotypes in image search results
  for occupations.
\newblock In {\em Proceedings of the 33rd Annual ACM Conference on Human
  Factors in Computing Systems}, pages 3819--3828. ACM, 2015.

\bibitem{kearns_preventing_2017}
Michael Kearns, Seth Neel, Aaron Roth, and Zhiwei~Steven Wu.
\newblock Preventing fairness gerrymandering: {{Auditing}} and learning for
  subgroup fairness.
\newblock {\em arXiv preprint arXiv:1711.05144}, 2017.

\bibitem{kennedy_improving_2017-1}
Ryan Kennedy, Stefan Wojcik, and David Lazer.
\newblock Improving election prediction internationally.
\newblock {\em Science}, 355(6324):515--520, 2017.

\bibitem{kerr_harking_1998}
Norbert~L. Kerr.
\newblock {{HARKing}}: {{Hypothesizing}} after the results are known.
\newblock {\em Personality and Social Psychology Review}, 2(3):196--217, 1998.

\bibitem{kleinberg2018inherent}
Jon Kleinberg.
\newblock Inherent trade-offs in algorithmic fairness.
\newblock In {\em ACM SIGMETRICS Performance Evaluation Review}, volume~46,
  pages 40--40. ACM, 2018.

\bibitem{krippendorff2004reliability}
Klaus Krippendorff.
\newblock Reliability in content analysis.
\newblock {\em Human communication research}, 30(3):411--433, 2004.

\bibitem{larson_how_2016}
Jeff Larson and Julia Angwin.
\newblock How {{We Analyzed}} the {{COMPAS Recidivism Algorithm}}.
\newblock {\em ProPublica}, May 2016.

\bibitem{lazer_parable_2014}
David Lazer, Ryan Kennedy, Gary King, and Alessandro Vespignani.
\newblock The parable of {{Google}} flu: Traps in big data analysis.
\newblock {\em Science}, 343(6176):1203--1205, 2014.

\bibitem{lazer_computational_2009-1}
David Lazer, Alex~Sandy Pentland, Lada Adamic, Sinan Aral, Albert~Laszlo
  Barabasi, Devon Brewer, Nicholas Christakis, Noshir Contractor, James Fowler,
  Myron Gutmann, et~al.
\newblock Computational social science.
\newblock {\em Science (New York, NY)}, 323(5915):721, 2009.

\bibitem{lazer_data_2017}
David Lazer and Jason Radford.
\newblock Data ex {{Machina}}: {{Introduction}} to {{Big Data}}.
\newblock {\em Annual Review of Sociology}, 43(1):19--39, 2017.

\bibitem{levendusky2009partisan}
Matthew Levendusky.
\newblock {\em The partisan sort: How liberals became Democrats and
  conservatives became Republicans}.
\newblock University of Chicago Press, 2009.

\bibitem{li_aoc_2019}
Danny Li.
\newblock {{AOC Is Right}}: {{Algorithms Will Always Be Biased As Long As
  There}}'s {{Systemic Racism}} in {{This Country}}.
\newblock
  https://slate.com/news-and-politics/2019/02/aoc-algorithms-racist-bias.html,
  February 2019.

\bibitem{lipton_mythos_2016}
Zachary~C. Lipton.
\newblock The mythos of model interpretability.
\newblock {\em arXiv preprint arXiv:1606.03490}, 2016.

\bibitem{liu_topic-link_2009}
Yan Liu, Alexandru {Niculescu-Mizil}, and Wojciech Gryc.
\newblock Topic-link {{LDA}}: Joint models of topic and author community.
\newblock In {\em Proceedings of the 26th {{Annual International Conference}}
  on {{Machine Learning}}}, pages 665--672. {ACM}, 2009.

\bibitem{lucas_computer-assisted_2015}
Christopher Lucas, Richard~A. Nielsen, Margaret~E. Roberts, Brandon~M. Stewart,
  Alex Storer, and Dustin Tingley.
\newblock Computer-{{Assisted Text Analysis}} for {{Comparative Politics}}.
\newblock {\em Political Analysis}, 23(2):254--277, 2015/ed.

\bibitem{lui2012langid}
Marco Lui and Timothy Baldwin.
\newblock langid. py: An off-the-shelf language identification tool.
\newblock In {\em Proceedings of the ACL 2012 system demonstrations}, pages
  25--30. Association for Computational Linguistics, 2012.

\bibitem{lundberg2019privacy}
Ian Lundberg, Arvind Narayanan, Karen Levy, and Matthew~J Salganik.
\newblock Privacy, ethics, and data access: A case study of the fragile
  families challenge.
\newblock {\em Socius}, 5:2378023118813023, 2019.

\bibitem{marsden1993network}
Peter~V Marsden and Noah~E Friedkin.
\newblock Network studies of social influence.
\newblock {\em Sociological Methods \& Research}, 22(1):127--151, 1993.

\bibitem{martin_egg_1991}
Emily Martin.
\newblock The {{Egg}} and the {{Sperm}}: {{How Science Has Constructed}} a
  {{Romance Based}} on {{Stereotypical Male}}-{{Female Roles}}.
\newblock {\em Signs: Journal of Women in Culture and Society}, 16(3):485--501,
  April 1991.

\bibitem{mason_i_2015}
Lilliana Mason.
\newblock "{{I Disrespectfully Agree}}": {{The Differential Effects}} of
  {{Partisan Sorting}} on {{Social}} and {{Issue Polarization}}.
\newblock {\em American Journal of Political Science}, 59(1):128--145, 2015.

\bibitem{mcpherson_birds_2001}
M.~McPherson, Lynn {Smith-Lovin}, and J.~Cook.
\newblock Birds of a {{Feather}}: {{Homophily}} in {{Social Networks}}.
\newblock {\em Annual Review of Sociology}, (1):415--444, 2001.

\bibitem{mitchell_prediction-based_2018}
Shira Mitchell, Eric Potash, and Solon Barocas.
\newblock Prediction-{{Based Decisions}} and {{Fairness}}: {{A Catalogue}} of
  {{Choices}}, {{Assumptions}}, and {{Definitions}}.
\newblock {\em arXiv:1811.07867 [stat]}, November 2018.

\bibitem{morstatter_search_2017}
Fred Morstatter and Huan Liu.
\newblock In search of coherence and consensus: Measuring the interpretability
  of statistical topics.
\newblock {\em The Journal of Machine Learning Research}, 18(1):6177--6208,
  2017.

\bibitem{MukherjeeJointAuthorSentiment2014}
S~Mukherjee.
\newblock Joint {{Author Sentiment Topic Model}}.
\newblock In {\em {{SDM}}}, 2014.

\bibitem{nelson_computational_2017}
Laura~K. Nelson.
\newblock Computational grounded theory: {{A}} methodological framework.
\newblock {\em Sociological Methods \& Research}, page 0049124117729703, 2017.

\bibitem{oconnor_computational_2011}
B.~O'Connor, D.~Bamman, and N.~A. Smith.
\newblock Computational {{Text Analysis}} for {{Social Science}}: {{Model
  Assumptions}} and {{Complexity}}.
\newblock {\em public health}, 41(42):43, 2011.

\bibitem{olteanu_social_2016}
Alexandra Olteanu, Carlos Castillo, Fernando Diaz, and Emre Kiciman.
\newblock Social {{Data}}: {{Biases}}, {{Methodological Pitfalls}}, and
  {{Ethical Boundaries}}.
\newblock {{SSRN Scholarly Paper}} ID 2886526, {Social Science Research
  Network}, Rochester, NY, December 2016.

\bibitem{omi2014racial}
Michael Omi and Howard Winant.
\newblock {\em Racial formation in the United States}.
\newblock Routledge, 2014.

\bibitem{passonneau_benefits_2014}
Rebecca~J. Passonneau and Bob Carpenter.
\newblock The benefits of a model of annotation.
\newblock {\em Transactions of the Association for Computational Linguistics},
  2:311--326, 2014.

\bibitem{pearl_seven_2019}
Judea Pearl.
\newblock The seven tools of causal inference, with reflections on machine
  learning.
\newblock {\em Communications of the ACM}, 62(3):54--60, February 2019.

\bibitem{poole1990patterns}
Keith~T Poole and Howard Rosenthal.
\newblock Patterns of congressional voting.
\newblock 1990.

\bibitem{raykar_learning_2010}
Vikas~C. Raykar, Shipeng Yu, Linda~H. Zhao, Gerardo~Hermosillo Valadez, Charles
  Florin, Luca Bogoni, and Linda Moy.
\newblock Learning from crowds.
\newblock {\em Journal of Machine Learning Research}, 11(Apr):1297--1322, 2010.

\bibitem{ribeiro_why_2016}
Marco~Tulio Ribeiro, Sameer Singh, and Carlos Guestrin.
\newblock "{{Why Should I Trust You}}?": {{Explaining}} the {{Predictions}} of
  {{Any Classifier}}.
\newblock {\em arXiv:1602.04938 [cs, stat]}, February 2016.

\bibitem{rickford1999african}
John~R Rickford and William Labov.
\newblock {\em African American vernacular English: Features, evolution,
  educational implications}.
\newblock Blackwell Malden, MA, 1999.

\bibitem{roberts_structural_2013-1}
Margaret~E. Roberts, Brandon~M. Stewart, Dustin Tingley, and Edoardo~M.
  Airoldi.
\newblock The structural topic model and applied social science.
\newblock In {\em Advances in Neural Information Processing Systems Workshop on
  Topic Models: Computation, Application, and Evaluation}, pages 1--20, 2013.

\bibitem{roberts_structural_2014}
Margaret~E. Roberts, Brandon~M. Stewart, Dustin Tingley, Christopher Lucas,
  Jetson {Leder-Luis}, Shana~Kushner Gadarian, Bethany Albertson, and David~G.
  Rand.
\newblock Structural topic models for open-ended survey responses.
\newblock {\em American Journal of Political Science}, 58(4):1064--1082, 2014.

\bibitem{rohrer_thinking_2018}
Julia~M. Rohrer.
\newblock Thinking {{Clearly About Correlations}} and {{Causation}}:
  {{Graphical Causal Models}} for {{Observational Data}}.
\newblock {\em Advances in Methods and Practices in Psychological Science},
  1(1):27--42, March 2018.

\bibitem{rosen2004author}
Michal Rosen-Zvi, Thomas Griffiths, Mark Steyvers, and Padhraic Smyth.
\newblock The author-topic model for authors and documents.
\newblock In {\em Proceedings of the 20th conference on Uncertainty in
  artificial intelligence}, pages 487--494. AUAI Press, 2004.

\bibitem{salganik2019introduction}
Matthew~J Salganik, Ian Lundberg, Alexander~T Kindel, and Sara McLanahan.
\newblock Introduction to the special collection on the fragile families
  challenge.
\newblock {\em Socius}, 5:2378023119871580, 2019.

\bibitem{SchwartzPersonalityGenderAge2013a}
H.~Andrew Schwartz, Johannes~C. Eichstaedt, Margaret~L. Kern, Lukasz
  Dziurzynski, Stephanie~M. Ramones, Megha Agrawal, Achal Shah, Michal
  Kosinski, David Stillwell, Martin E.~P. Seligman, and Lyle~H. Ungar.
\newblock Personality, {{Gender}}, and {{Age}} in the {{Language}} of {{Social
  Media}}: {{The Open}}-{{Vocabulary Approach}}.
\newblock {\em PLoS ONE}, 8(9):e73791, September 2013.

\bibitem{selbst_fairness_2018}
Andrew~D. Selbst, Danah Boyd, Sorelle Friedler, Suresh Venkatasubramanian, and
  Janet Vertesi.
\newblock Fairness and {{Abstraction}} in {{Sociotechnical Systems}}.
\newblock {{SSRN Scholarly Paper}} ID 3265913, {Social Science Research
  Network}, {Rochester, NY}, August 2018.

\bibitem{small_someone_2017}
Mario~Luis Small.
\newblock {\em Someone to {{Talk}} To}.
\newblock {Oxford University Press}, 2017.

\bibitem{smith-lovin_strength_2007-1}
Lynn {Smith-Lovin}.
\newblock The {{Strength}} of {{Weak Identities}}: {{Social Structural
  Sources}} of {{Self}}, {{Situation}} and {{Emotional Experience}}.
\newblock {\em Social Psychology Quarterly}, 70(2):106--124, June 2007.

\bibitem{snow_cheap_2008}
Rion Snow, Brendan O'Connor, Daniel Jurafsky, and Andrew~Y. Ng.
\newblock Cheap and fast\textemdash{}but is it good?: Evaluating non-expert
  annotations for natural language tasks.
\newblock In {\em Proceedings of the Conference on Empirical Methods in Natural
  Language Processing}, pages 254--263. {Association for Computational
  Linguistics}, 2008.

\bibitem{szegedy2017inception}
Christian Szegedy, Sergey Ioffe, Vincent Vanhoucke, and Alexander~A Alemi.
\newblock Inception-v4, inception-resnet and the impact of residual connections
  on learning.
\newblock In {\em Thirty-First AAAI Conference on Artificial Intelligence},
  2017.

\bibitem{tavory2014abductive}
Iddo Tavory and Stefan Timmermans.
\newblock {\em Abductive analysis: Theorizing qualitative research}.
\newblock University of Chicago Press, 2014.

\bibitem{todorov2008understanding}
Alexander Todorov, Chris~P Said, Andrew~D Engell, and Nikolaas~N Oosterhof.
\newblock Understanding evaluation of faces on social dimensions.
\newblock {\em Trends in cognitive sciences}, 12(12):455--460, 2008.

\bibitem{toole_jameson_l._tracking_2015}
{Toole Jameson L.}, {Lin Yu-Ru}, {Muehlegger Erich}, {Shoag Daniel},
  {Gonz{\'a}lez Marta C.}, and {Lazer David}.
\newblock Tracking employment shocks using mobile phone data.
\newblock {\em Journal of The Royal Society Interface}, 12(107):20150185, June
  2015.

\bibitem{tsur_frame_2015}
Oren Tsur, Dan Calacci, and David Lazer.
\newblock A {{Frame}} of {{Mind}}: {{Using Statistical Models}} for
  {{Detection}} of {{Framing}} and {{Agenda Setting Campaigns}}.
\newblock In {\em {{ACL}} (1)}, pages 1629--1638, 2015.

\bibitem{tufekci_big_2014}
Zeynep Tufekci.
\newblock Big {{Questions}} for {{Social Media Big Data}}:
  {{Representativeness}}, {{Validity}} and {{Other Methodological Pitfalls}}.
\newblock In {\em {{ICWSM}} '14: {{Proceedings}} of the 8th {{International
  AAAI Conference}} on {{Weblogs}} and {{Social Media}}.}, 2014.

\bibitem{tufekci_big_2014-1}
Zeynep Tufekci.
\newblock Big {{Questions}} for {{Social Media Big Data}}:
  {{Representativeness}}, {{Validity}} and {{Other Methodological Pitfalls}}.
\newblock In {\em {{ICWSM}} '14: {{Proceedings}} of the 8th {{International
  AAAI Conference}} on {{Weblogs}} and {{Social Media}}.}, 2014.

\bibitem{van_bavel_partisan_2018}
Jay~J. Van~Bavel and Andrea Pereira.
\newblock The {{Partisan Brain}}: {{An Identity}}-{{Based Model}} of
  {{Political Belief}}.
\newblock {\em Trends in Cognitive Sciences}, 22(3):213--224, March 2018.

\bibitem{vaswani2017attention}
Ashish Vaswani, Noam Shazeer, Niki Parmar, Jakob Uszkoreit, Llion Jones,
  Aidan~N Gomez, {\L}ukasz Kaiser, and Illia Polosukhin.
\newblock Attention is all you need.
\newblock In {\em Advances in neural information processing systems}, pages
  5998--6008, 2017.

\bibitem{wallace_universal_2019}
Eric Wallace, Shi Feng, Nikhil Kandpal, Matt Gardner, and Sameer Singh.
\newblock Universal adversarial triggers for nlp.
\newblock {\em arXiv preprint arXiv:1908.07125}, 2019.

\bibitem{wallach_computational_2018}
Hanna Wallach.
\newblock Computational social science {$\not =$} computer science + social
  data.
\newblock {\em Communications of the ACM}, 61(3):42--44, February 2018.

\bibitem{wang_forecasting_2015}
Wei Wang, David Rothschild, Sharad Goel, and Andrew Gelman.
\newblock Forecasting elections with non-representative polls.
\newblock {\em International Journal of Forecasting}, 31(3):980--991, 2015.

\bibitem{wang_deep_2018}
Yilun Wang and Michal Kosinski.
\newblock Deep neural networks are more accurate than humans at detecting
  sexual orientation from facial images.
\newblock {\em Journal of personality and social psychology}, 114(2):246, 2018.

\bibitem{waseem_are_2016}
Zeerak Waseem.
\newblock Are {{You}} a {{Racist}} or {{Am I Seeing Things}}? {{Annotator
  Influence}} on {{Hate Speech Detection}} on {{Twitter}}.
\newblock {\em NLP+ CSS 2016}, page 138, 2016.

\bibitem{watts_everything_2011}
Duncan~J. Watts.
\newblock {\em Everything Is Obvious:* {{Once}} You Know the Answer}.
\newblock {Crown Business}, 2011.

\bibitem{wu_automated_2016}
Xiaolin Wu and Xi~Zhang.
\newblock Automated inference on criminality using face images.
\newblock {\em arXiv preprint arXiv:1611.04135}, pages 4038--4052, 2016.

\bibitem{yan2013biterm}
Xiaohui Yan, Jiafeng Guo, Yanyan Lan, and Xueqi Cheng.
\newblock A biterm topic model for short texts.
\newblock In {\em Proceedings of the 22nd international conference on World
  Wide Web}, pages 1445--1456. ACM, 2013.

\bibitem{zagoruyko_paying_2016}
Sergey Zagoruyko and Nikos Komodakis.
\newblock Paying more attention to attention: {{Improving}} the performance of
  convolutional neural networks via attention transfer.
\newblock {\em arXiv preprint arXiv:1612.03928}, 2016.

\bibitem{zuberi2008white}
Tukufu Zuberi, Eduardo Bonilla-Silva, et~al.
\newblock {\em White logic, white methods: Racism and methodology}.
\newblock Rowman \& Littlefield Publishers, 2008.

\end{thebibliography}
\end{document}